\documentclass[%
 aip,
 jap,
 amsmath,amssymb,
 reprint,%
]{revtex4-1}

\usepackage{graphicx}
\usepackage{dcolumn}
\usepackage{bm}
\usepackage{times}
\DeclareGraphicsExtensions{.eps}
\newcolumntype{C}{>{\centering\arraybackslash}X}
\usepackage{epstopdf}
\usepackage{lipsum}
\usepackage{tabularx}
\usepackage{booktabs}
\usepackage{array}
\usepackage{natbib}
\usepackage{amsmath}
\usepackage{color}
\usepackage{epstopdf}
\usepackage{grffile}
\usepackage{caption}
\usepackage{boxhandler}
\usepackage{float}
\usepackage{lipsum}
\begin{document}

\preprint{AIP/123-QED}

\title{Prospects of Zero Schottky Barrier Height in a Graphene Inserted MoS$_2$-Metal Interface  }

\author{Anuja Chanana and Santanu Mahapatra}

\affiliation{ Nano-Scale Device Research Laboratory, Department of Electronic Systems Engineering, Indian Institute of Science (IISc) Bangalore, Bangalore 560012, India
}%
\date{\today}

\begin{abstract}
A low Schottky barrier height (SBH) at source/drain contact is essential for achieving high drive current in atomic layer MoS$_2$ channel based field-effect transistors. Approaches such as choosing metals with appropriate work functions and chemical doping are employed previously to improve the carrier injection from the contact electrodes to the channel and to mitigate the SBH between the MoS$_2$ and metal. Recent experiments demonstrate significant SBH reduction when graphene layer is inserted between metal slab (Ti and Ni) and MoS$_2$. However, the physical or chemical origin of this phenomenon is not yet clearly understood. In this work, density functional theory (DFT) simulations are performed, employing pseudopotentials with very high basis sets to get insights of the charge transfer between metal and monolayer MoS$_2$ through the inserted graphene layer. Our atomistic simulations on 16 different interfaces involving five different metals (Ti, Ag, Ru, Au and Pt)  reveal that: (i) such a decrease in SBH is not consistent among various metals, rather an increase in SBH is observed in case of Au and Pt (ii) unlike MoS$_2$-metal interface, the projected dispersion of MoS$_2$ remains preserved in any MoS$_2$-graphene-metal system with shift in the bands on the energy axis. (iii) a proper choice of metal (e.g., Ru) may exhibit ohmic nature in a graphene inserted MoS$_2$-metal contact.  These understandings would provide a direction in developing high performance transistors involving hetero atomic layers as contact electrodes.
\end{abstract}

\keywords{Graphene, MoS$_2$, metal contact, Schottky Barrier Height, Electron Density Difference, Density Functional Theory} 
\maketitle

\section{\label{sec:level1}Introduction}

Since the first demonstration of monolayer MoS$_2$ channel based metal oxide semiconductor field effect transistor (MOSFET) by the EPFL research team \cite{radisavljevic2011single}, the nanoelectronics community has shown tremendous interest towards 2D layered-materials \cite{}. These materials promise to offer exceptional electrostatic integrity and therefore are suitable for decananometer technology nodes\cite{ITRS}. However, the experimental reports of the drain current for such atomic layer channel based MOSFETs are much lower than the desired ON current value required for technology downscaling. One of the primary reasons for such low ON current is inefficient carrier injection from the source to the channel, which originates from the significant Schottky  barrier height (SBH) formed between the 2D channel material and the metal electrode (by SBH we mean n-SBH unless it is specified elsewhere). Obtaining very low or even zero SBH at source/drain contacts is one of the most essential and challenging tasks for realizing high performance atomically thin material based MOSFETs. Attempts are made to reduce SBH by choosing low work function metals (e.g. Scandium \cite{das2012high}, Molybdenum \cite{:/content/aip/journal/apl/104/9/10.1063/1.4866340} etc.) or even by employing low pressure metal deposition techniques \cite{english2014improving}. Novel doping methodology for TMD's\cite{doi:10.1021/nl502603d, doi:10.1021/nl503251h, :/content/aip/journal/apl/104/9/10.1063/1.4867197, doi:10.1021/nl301702r} is also proposed to reduce the SBH. Very recently, it is demonstrated experimentally that by inserting graphene layer between  MoS$_2$ and metal electrode (Ti\cite{du2014field} and Ni \cite{leong2014low}), SBH can be reduced significantly and hence greatly improve the drive current of the device. However, a detailed theoretical understanding of the underlying mechanism of such SBH reduction phenomena by inserting graphene layer is still lacking. It is also not clear if such technique successfully reduces SBH for the other metals commonly used as contact electrodes.

We utilize the density function theory (DFT) simulations to analyse the contact nature of the interfaces formed between monolayer MoS$_2$ and graphene-metal heterocontacts. The study is conducted for 5 different metals (Ti, Ag, Ru, Au and Pt) which are commonly used in experiments and the work function (WF) spans from low (Ti) to high (Pt) with an average interval of 0.25 eV. Both chemisorption and physiosorption interface metal surfaces with graphene are taken into account to develop better perception of the problem. We first simulate the MoS$_2$-graphene and graphene-metal systems separately and analyse their electronic structures. These understandings are then used to analyse the simulated characteristics of complex MoS$_2$-graphene-metal interface. To compare the SBH of a graphene inserted systems, the individual MoS$_2$-metal interfaces are also studied.  A thorough examination of 16 different interface structures shows that SBH reduction through graphene insertion in a metal-MoS$_2$ contact is not always obtained for different metals. While we observe such reduction for Ti (in agreement with experiment), Ru and Ag; an increase in SBH is observed in case of Au and Pt. It is further demonstrated that SBH in MoS$_2$-graphene-metal structure is governed by the property of graphene adsorbed metal surface by analyzing the Projected Density of States (PDOS). The graphene insertion in a MoS$_2$-metal contact preserves the dispersion nature of the MoS$_2$ despite of graphene-metal interface nature. Finally we show by electron density difference (EDD) investigation that choice of appropriate metal (as happens for Ru) may help to obtain pure ohmic contact in a MoS$_2$-graphene-metal system. It is worth noting that recent DFT studies on monolayer Boron Ntirde inserted MoS$_2$-metal contact also reveals zero Schottkey barrier nature with Co and Ni\cite{farmanbar2015controlling}.

\section{\label{sec:level1}Computational Details and Methods}
DFT code as implemented in Atomistix Tool Kit \cite{QumWS} employing Local Density Approximation (LDA) with Perdew-Zunger parametrization (PZ) \cite{PhysRevB.23.5048} as the exchange correlation functional is used for the present study. We first calculate the band gap of monolayer MoS$_2$ with the lattice parameter 3.1604 \hspace{0.1 cm}\AA \hspace{0.1cm} and found it to be 1.8 eV, which is consistent with the experimental studies\cite{PhysRevLett.105.136805}. Pseudopotentials conceptualized using the fully relativistic all-electron calculation \cite{PhysRevB.58.3641} as developed by  Hartwingster-Goedecker-Hutter (HGH) with Tier 8 basis set are adopted for each element. The Tier 8 basis set in ATK includes maximum number of atomic orbital contributions for HGH pseudopotential. We use such a higher basis set so that the dispersion of graphene-metal (especially graphene-gold\cite{slawinska2011doping}) and MoS$_2$-graphene interfaces \cite{C1NR10577A} are persistent with the previous reports and thus assures the accuracy in the dispersion of complex MoS$_2$-graphene-metal systems. The iteration steps are set as 100 using Pulay mixer algorithm as the iteration control parameter with a tolerance value upto 10$^{-5}$ Hartree. The Poisson solver we followed is fast Fourier transform (FFT). Density mesh cut off of 75 Hartree and a k point sampling of 9x9x1 under Monkhorst Pack scheme for the Broiullin Zone are chosen for the simulations. All the unit cells are relaxed using limited memory Broyden Fletcher Goldfarb Shannon method \cite{LBFGS} until the forces on the atom are 0.01 eV/\AA. 

\section{\label{sec:level1}Results and Discussions}

\begin{figure*}[!t]
\includegraphics[width =1.6\columnwidth]{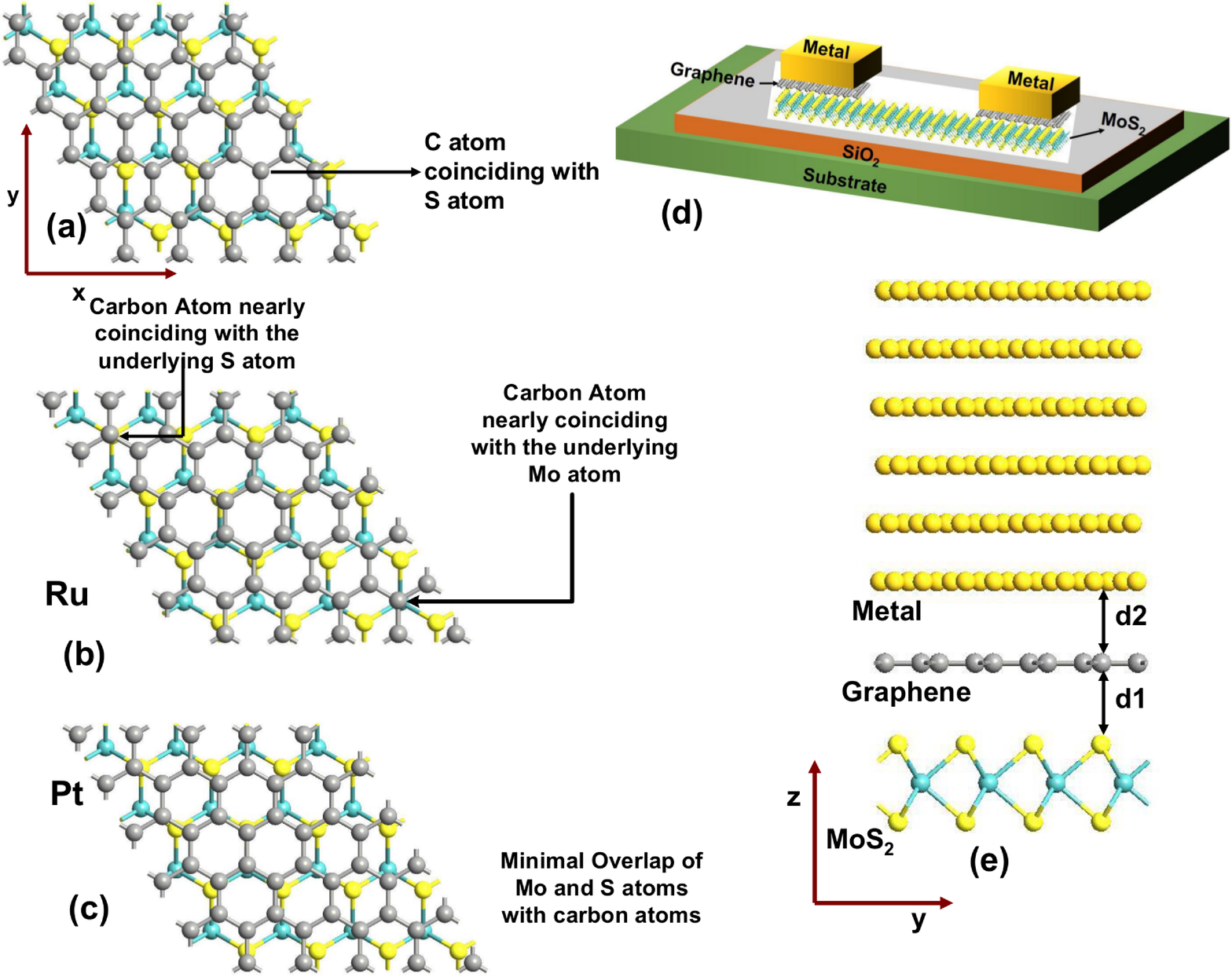}
\caption{(a) Top view of graphene (5x5 supercell) on MoS$_2$ (4x4 supercell) with equilibrium interlayer distance in a TS configuration (C atom on top of S atom of MoS$_2$) . The grey, blue and yellow balls indicate carbon, molybdenum and sulfur atoms respectively. Variation in the overlap of Mo and S atoms of monolayer MoS$_2$ with carbon atoms of monolayer graphene for (b) MoS$_2$-graphene-Ru and (c) MoS$_2$-graphene-Pt.(d) Transistor schematic using graphene-metal heterocontact interfaced with monolayer MoS$_2$ (e) Side view of graphene inserted MoS$_2$-Au contact with interlayer distance (z-direction) corresponding to individual interface structures. The metal surface is cleaved along $<$111$>$ and 6 atomic layers are used to make the interface. The atomic configurations are represented by Ball-Stick model.}
\captionsetup{format=hang}
\label{Figure 1}
\end{figure*}

\subsection{Interface Geometry}
Keeping in mind the commensurability condition, the interface is formed by 4x4 MoS$_2$ supercell (lattice parameter = 12.6416 \AA ) and 5x5 graphene supercell (lattice parameter = 12.306 \AA) and the mean strain on graphene is found to be 1.8\% which is in close agreement with earlier reports\cite{C1NR10577A}. In the resultant structure comprising of graphene and MoS$_2$ it is observed that one S atom coincides with the carbon atom in graphene respectively forming TS configuration (C atom on top of S atom of MoS$_2$) as shown in Fig.1(a) (Top View). It is worth noting that when the MoS$_2$ supercell is matched with graphene-metal heterocontact, the TS configuration between MoS$_2$ and graphene is not retained. This happens because the graphene is already interface matched with the metal slab and the atomic positions of carbon change with respect to  metal. Hence the MoS$_2$ atomic positions vary in accordance with the graphene-metal interface which may be physiosorption or chemisorption. Figure 1 (b) and (c) feature the overlap of carbon atom of graphene and the molybdenum and sulfur atoms of MoS$_2$ in a complex graphene inserted MoS$_2$-metal interface for Ru (chemisorption) and Pt(physiosorption). For Ru there is a close overlap of carbon atom with the underlying Mo and S atoms shown by black arrows. Both the Mo and S atom are located at the periphery of the hexagonal lattice. However, the overlap vanishes in case of Pt and there is no exact coincidence in the atomic positions for both the layers. This results due to variations in atomic positions of MoS$_2$ supercell w.r.t to graphene adsorbed metal interface and is different for different metals. The lattice of MoS$_2$ and graphene comes to a close equalization for 7x7 MoS$_2$ supercell (lattice parameter = 22.1228 \AA) and 9x9 graphene(lattice parameter = 22.1508 \AA) with a mean strain of 0.084\%. To save the computational cost, for MoS$_2$-graphene and MoS$_2$-graphene-metal interface, we continued with the previous lattice parameter for the present analysis. $<$111$>$ cleaved surface for Au, Ag and Pt and $<$0001$>$ cleaved surface for Ti and Ru each with six layers are interfaced and strained to match the supercells formed with monolayer graphene supercell and monolyer MoS$_2$ supercell and hetero MoS$_2$-graphene interface. \\

\begin{table*}[htb]
\centering
\caption{\label{}Strain applied in all the interfaces, Calculated equilibrium distances (z-direction) corresponding to minimum binding energy(BE), BE values, Dirac Cone shift only applicable to physisorped interfaces involving graphene, Schottky Barrier Heights(p-type and n-type) corresponding to interfaces with MoS$_2$, Band gap (Eg) values for MoS$_2$-metal interface and MoS$_2$-graphene-metal interface calculated by adding p-type SBH and n-type SBH}

\begin{ruledtabular}
\begin{tabular*}{\hsize}{l*{16}{p{0.05\hsize}}}
System & MoS$_2$-G & G-Au & G-Pt & G-Ti & G-Ag & G-Ru & MoS$_2$-Au & MoS$_2$-Pt & MoS$_2$-Ti & MoS$_2$-Ag & MoS$_2$-Ru & MoS$_2$-G-Au & MoS$_2$-G-Pt & MoS$_2$-G-Ti & MoS$_2$-G-Ag & MoS$_2$-G-Ru\\
&  &  &  &  &  &  &  &  &  &  &  &  &  &  &  & \\
Strain(\%) & 1.8 & 1.8 & 1.2 & 2.8 & 1.5 & 0.5 & 0.38 & 0.38 & 1.1 & 0.26 & 1.3 & 1.8 & 1.8 & 1.8 & 1.8 & 1.8\\
&  &  &  &  &  &  &  &  &  &  &  &  &  &  &  & \\
d(\AA) & 3.3 & 3.3 & 3.2 & 2.1 & 3.2 & 2.2 & 2.7 & 2.3 & 2.2 & 2.5 & 2.2 & d1=3.3 & d1=3.3 & d1=3.3 & d1=3.3 & d1=3.3\\
&  &  &  &  &  &  &  &  &  &  &  & d2=3.3 & d2=3.2 & d2=2.1 & d2=3.2 & d2=2.2\\
&  &  &  &  &  &  &  &  &  &  &  &  &  &  &  & \\
BE(eV) & -1.86 & -2.5 & -3 & -19 & -2.15 & -5 & -5.9 & -9.55 & -17.5 & -6.5 & -14.8 & -4.5 & -5.16 & -23.81 & -4.3 & -12.43\\
&  &  &  &  &  &  &  &  &  &  &  &  &  &  &  & \\
${\Delta}${E$_F$} & -0.02 & 0.1 & 0.28 & --- & -0.423 & --- & --- & --- & --- & --- & --- & -0.106 & 0.127 & --- & -0.057 & ---\\
&  &  &  &  &  &  &  &  &  &  &  &  &  &  &  & \\
n-SBH & 0.65 & --- & --- & --- & --- & --- & 0.64 & 0.81 & 0.382 & 0.373 & 0.56 & 0.663 & 0.916 & 0.26 & 0.25 & 0.018\\
 &  &  &  &  &  &  &  &  &  &  &  &  &  &  &  & \\
p-SBH & 1.14 & --- & --- & --- & --- & --- & 1.2 & 1.09 & 1.67  & 1.48 & 1.34  & 1.14 & 0.89 & 1.55  & 1.56  & 1.79 \\
 &  &  &  &  &  &  &  &  &  &  &  &  &  &  &  & \\
E$_g$ (eV) & 1.79 & --- & --- & --- & --- & --- & 1.84 & 1.9 & 2.052 & 1.853 & 1.9 & 1.803 & 1.806 & 1.81 & 1.81 & 1.808\\
 &  &  &  &  &  &  &  &  &  &  &  &  &  &  &  & \\
\end{tabular*}
\end{ruledtabular}
\end{table*}


Figure 1 (d) shows the transistor schematic where the graphene layer is sandwiched between MoS$_2$ and metal for a top contact geometry used for the current study. The hybrid structure showing supercell formed using graphene-gold heterocontact and monolayer MoS$_2$ is presented in Figure 1 (e). Table I provides the interface strain, equilibrium interlayer distance (d1 and d2), Binding Energies (BE), Dirac Cone Shift(${\Delta}${E$_F$}) and the respective SBH for all the systems simulated in this work. The BE for MoS$_2$-metal and graphene-metal system is defined as following BE (MoS$_2$/graphene  -  metal) = TE (MoS$_2$/graphene + metal) - TE (metal)- TE (MoS$_2$/graphene) and for complex MoS$_2$-graphene-metal system as BE(MoS$_2$-graphene-metal) = TE (MoS$_2$-graphene + metal) - TE (metal)- TE (MoS$_2$)- TE (graphene).

For a graphene inserted MoS$_2$-metal interface, the distance between monolayer graphene-monolayer MoS$_2$ supercell (d1) and graphene-metal system (d2) is kept same as the one obtained for individual interfaces. A distance of more than 20 \AA \hspace{0.1 cm} is adopted in a perpendicular direction normal to the interface to isolate the slab from false interactions between periodic structures. Since both the 2D materials used have a hexagonal lattice parameter, so we have maintained the hexagonal lattice geometry for all the interface structures.

We first conduct DFT simulations on simple graphene interfaces such as MoS$_2$-graphene and graphene metal to study their dispersion natures. Further based on these characteristics, the electronic properties of complex MoS$_2$-graphene-metal systems are analyzed. The values of (${\Delta}${E$_F$}) show that two metals (Ti and Ru) are chemisorped and 3 metals (Au,Pt and Ag) are physiosorped with both graphene and MoS$_2$-graphene heterostructure. Apart from Pt which is chemisorped with MoS$_2$, rest of the metals show an interface nature with MoS$_2$ similar to graphene . The BE values confirm the kind of nature whether chemisorption or physiosorption for graphene, MoS$_2$, MoS$_2$-graphene when interfaced with metals. For chemisorpotion interface, the values of BE are more negative in comparison to the physiosorped interface. From n-SBH values of MoS$_2$-metal and MoS$_2$-graphene-metal we see that there is an increase in n-SBH for Au and Pt while a decrease is observed for the rest. The values of (${\Delta}${E$_F$}) and SBH shows minor changes in a MoS$_2$-graphene-Au w.r.t graphene-Au and MoS$_2$-Au systems, thus confirming that graphene insertion in MoS$_2$-metal contact does not always ensure a SBH reduction. The same result is verified for Pt, where we see a significant increase of SBH value and shift in ${\Delta}${E$_F$} with graphene insertion w.r.t MoS$_2$-Pt and graphene-Pt system. An increase in n-SBH implies a decrease of p-SBH and it is more pronounced for Pt. While the other metals (Ag,Ti and Ru) show an SBH reduction with Ru exhibiting the maximum decrease.

\subsection{Electronic Structure Analysis}
As been observed in earlier reports\cite{chen2013tuning, gong2014unusual} for heterogeneous interfaces, obtaining the exact value of MoS$_2$ band gap and identification of VBM(valence band maxima) and CBM(conduction band minima) is difficult. Figure 2 shows the projected band structure and DOS of (a) MoS$_2$-Ru and (b) MoS$_2$-graphene-Ru interface. To determine the position of CBM and VBM in a MoS$_2$-metal interface, the projected band structure and PDOS of MoS$_2$ are kept alongside each other by aligning their Fermi level. The mid gap states in DOS are very high for MoS$_2$-Ru (chemisorption interface) as compared to MoS$_2$-graphene-Ru interface because graphene acts as a buffer layer between MoS$_2$ and Ru. In a MoS$_2$-Ru interface, the VBM position is apparently visible but the CBM position is ambiguous. The position of CBM is found out by measuring the band gap value from VBM position to an estimated CBM curvature where the value is closer to 1.8 eV. To confirm these positions DOS is placed beside and lines are drawn (black dotted lines), from CBM and VBM in band structure extending to the DOS region. In between this particular interval, the mid gap states contribution in DOS is minimal, which confirms the respective CBM and VBM positions. Depending on the type of interface (chemisorption and physiosorption), the amount of mid gap states vary in the particular band gap regime. The variation in the MoS$_2$ band gap is higher for chemisorped interfaces as compared to physiosorped interfaces. The same methodology is used for MoS$_2$-graphene-metal interface and is shown for Ru metal (b). The difference is that the VBM position is not very precise, so the VBM position w.r.t CBM is distinguished among various bands employing projected DOS. We see that the black dashed lines connecting the band edges in bandstructure with DOS connects perfectly at those energy levels where the contribution of PDOS is zero. Hence, in a graphene-inserted MoS$_2$-metal systems, the band gap value remains closer to the pristine MoS$_2$(1.8 eV). This shows that graphene layer acts as a perfect buffer between MoS$_2$ and metal and lessens the effect of metal on the band structure of MoS$_2$. In general, the chemisorption interfaces have higher mid gap states in comparison to physiosopred interface due to high amount of hybridization at the interface. This makes the determination of CBM and VBM edges become difficult and is seen for other metals (Pd and Ir) as well\cite{gong2014unusual}. The n-type SBH is calculated as E$_C$-E$_F$ and the p-type SBH is E$_F$-E$_V$ and is shown by black arrows along with the conduction and valence band edges.

\begin{figure}[!t]
\centering
\includegraphics[scale=0.4]{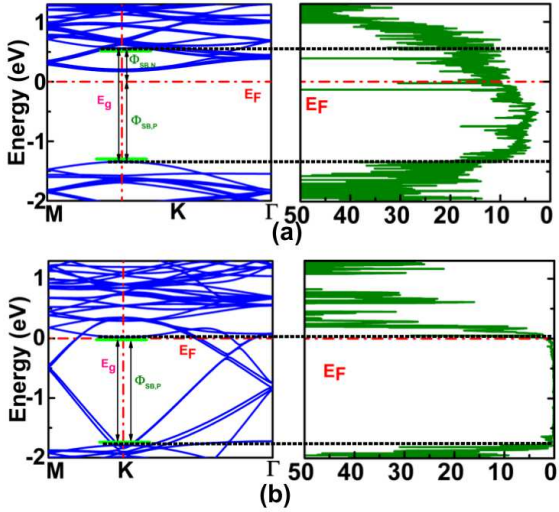}
\caption{Projected bandstructure and DOS of MoS$_2$ for (a) MoS$_2$-Ru interface and (b) MoS$_2$-graphene-Ru interface . Fermi level is positioned at zero and aligned to examine the positions of CBM and VBM.}   
\label{Figure 4}
\end{figure}

\begin{figure*}[!t]
\centering
\includegraphics[width =2.0\columnwidth]{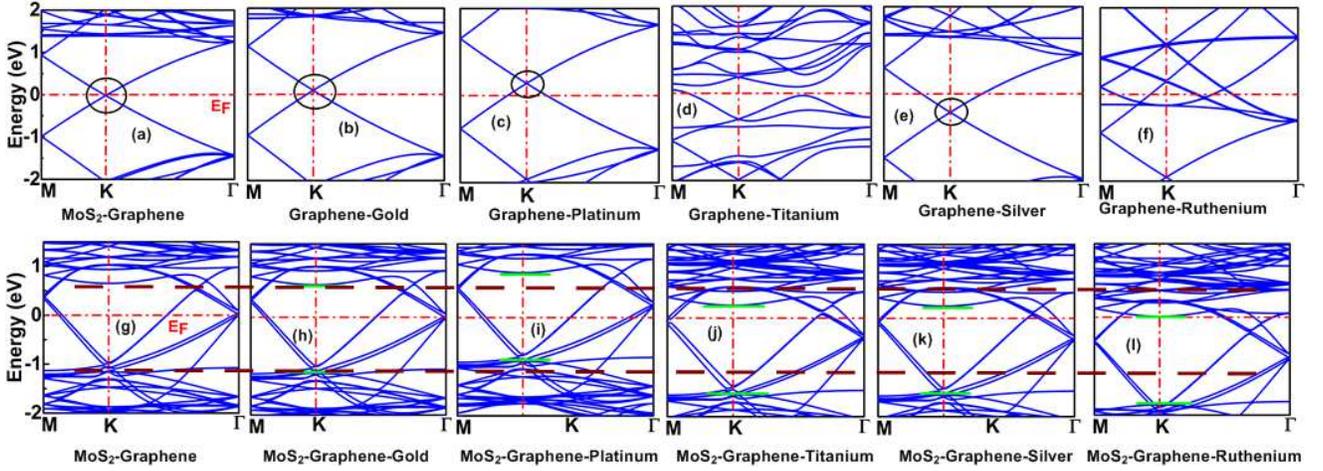}
\caption{Projected band structures of 5x5 graphene supercell for (a)MoS$_2$-Graphene, (b) Graphene-Gold (c) Graphene-Platinum (d) Graphene-Titanium (d) Graphene-Silver and (d) Graphene-Ruthenium interface. For physiosorption cases (a),(b), (c) and (e) the Dirac cone is retained and indicated by black circles. For chemisorption cases (d) and (f) the Dirac cone is completely vanished. Projected bandstructures of 4x4 MoS$_2$ supercell for (g) MoS$_2$-Graphene (h) MoS$_2$-Graphene-Au (i)MoS$_2$-Graphene-Pt (j)MoS$_2$-Graphene-Ti (k)MoS$_2$-Graphene-Ag and (l)MoS$_2$-Graphene-Ru interface. Since the midgap states are present in the heterogeneous interface the valence band maxima (VBM) and conduction band minima (CBM) are presented by green lines. The brown dashed lines are aligned with the CBM and VBM of projected MoS$_2$ of MoS$_2$-graphene interface in (g) and highlight the variation of respective CBM and VBM of other interface structures. The Fermi level is denoted by E$_F$ and is set as zero.}
\label{Figure 2}
\end{figure*}

\begin{figure*}[!t]
\centering
\includegraphics [width =2.0\columnwidth]{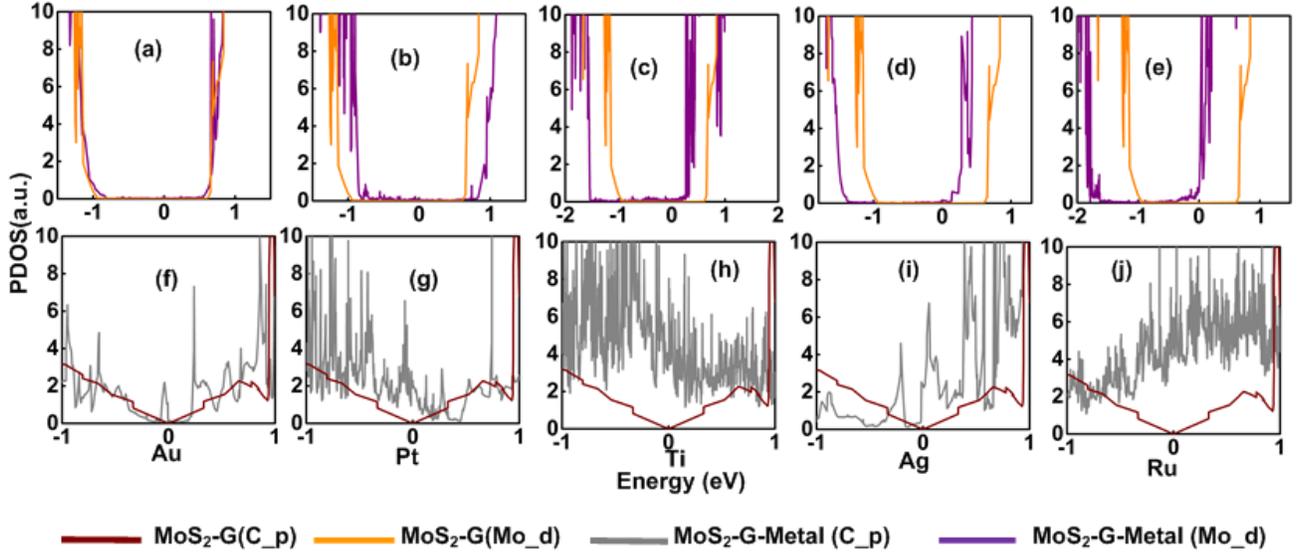}
\caption{Projected density of states of Mo-d orbital and C-p orbital MoS$_2$-graphene-metal heterocontacts interfaces for (a),(f)Au, (b),(g)Pt, (c),(h)Ti (d),(i)Ag and (e),(j)Ru systems superimposed with that MoS$_2$-graphene system. The legends are specified at bottom. The shifts in the orbitals of complex MoS$_2$-graphene-metal are compared w.r.t the MoS$_2$-graphene interface.}
\label{Figure 5}
\end{figure*}

Figure 3 show the projected band structure of carbon atoms for 5x5 graphene supercell for (a) MoS$_2$-graphene and (b)-(f) graphene-metal systems. The Dirac cone is preserved for only (a) MoS$_2$-graphene and for (b) graphene-gold (c) graphene-platinum and (e) graphene-silver interface with a shift w.r.t E$_F$ and is shown by black circles. This nature is completely lost for the chemisorption interfaces such as Ti and Ru. The shift of Dirac Cone in graphene-Ag is higher and opposite in nature when compared to both Pt and Au.  Figure 3(g)-(l) shows projected bandstructure of 4x4 MoS$_2$ supercell in a MoS$_2$-graphene interface (g) and graphene inserted MoS$_2$-metal interface (h)-(l). A brown dashed line is drawn to identify the relative shift in CBM and VBM of MoS$_2$-graphene-metal systems w.r.t the MoS$_2$-graphene system. The VBM and CBM are denoted by green lines. The CBM remains nearly same for MoS$_2$-graphene and MoS$_2$-graphene-Au systems and it lowers down for Ti and Ag, but substantial shift is observed for Ru where CBM moves to the proximity of Fermi Level. On the other hand, CBM shifts upwards for Pt. An Ohmic nature appears for Ru contact where we find n-type SBH to be almost zero (0.018 eV). The projected MoS$_2$ bandstucture nature of MoS$_2$-graphene is preserved for every MoS$_2$-graphene-metal interface and the interface states are found to be minimal. This again implies that graphene is successful in isolating the MoS$_2$ from metal with nearly equivalent zero midgap states and only shifts of CBM and VBM with respect to MoS$_2$-graphene system, and the graphene-metal interaction dictates the amount and nature of shift. 

\subsection{Density of States Analysis}
The relative shift of MoS$_2$ band edges and perturbation in the Dirac nature among various MoS$_2$-graphene-metal interfaces are highlighted in Figure 4 using the projected density of states (PDOS). Figure 4 (a)-(e) and (f)-(j) shows the PDOS of 4x4 MoS$_2$ supercell and 5x5 graphene supercell. We superimpose the PDOS of MoS$_2$-graphene system over the MoS$_2$-graphene-metal system to present the difference in amounts of hybridization for physiosorped and chemisorped metals. The p-orbital of carbon in graphene and d-orbital of Mo which are maximum contributors for the Dirac cone in graphene \cite{PhysRevB.79.195425} and  VBM, CBM positions in MoS$_2$\cite{gong2013band} respectively are used to study these effects.  In figure 4 (a)-(e), we see the relative shifts in the Mo-d orbital edges for various metals. They are consistent with the projected bandstructure shown in Figure 3 (g)-(l).In terms of hybridization, it is observed that the weakly chemisorped metal Pt (4(b)) and highly chemisorped metal Ti and Ru (4(c) and (e)), the CB and VB edges are highly perturbed in comparison to Au and Ag (4(a) and 4(d)).  From Figure 4(f) it is seen that Carbon p orbital contribution of MoS$_2$-graphene and MoS$_2$-graphene-Au nearly similar to each other since graphene is physiosorped on MoS$_2$ and Au and hence the Dirac nature is least perturbed. This nature of perturbation is observed in Ag and Pt as well and the Dirac cone gets shifted from the zero point Fermi level as compared to MoS$_2$-graphene system. When we compare the ${\Delta}${E$_F$} in graphene-metal with MoS$_2$-graphene-metal from Table 1, we see that the change in ${\Delta}${E$_F$} is highest for Ag when contacted with MoS$_2$-graphene system and hence the SBH reduction is also maximum for Ag for all physiosorped metals. The variation of Pt is higher as compared to Au since Pt has one electron less in d-orbital as compared to Au, so its more reactive. The nature of shift for Au and Pt also are opposite as compared to Ag and hence we see SBH reduction for Ag and SBH increment for both Au and Pt. For chemisorped metal Ti and Ru the Dirac nature is completely lost. The interlayer separation between graphene and Ti/Ru is very less in comparison to Au, Ag and Pt, hence we see higher interface states near the Fermi Level for Ti and Ru. This is indicative of strong and complex bonding between the carbon and metal atoms. Ti being the d-electron metal has a completely filled s-orbital thus is more reactive. It is chemisorped with graphene and MoS$_2$ so the perturbations in the orbitals are very high as compared to Au and reflects a complete distortion. Thus, we see SBH alterations for MoS$_2$-graphene-Ti w.r.t to MoS$_2$-graphene. In case of Ru both d and s orbitals are partially filled and the perturbations for Carbon-p lies intermediate between the two cases i.e. Au and Ti. Thus, it is expected that the its chemical reactivity also follows the same trend. But the change in SBH in Ru is higher as compared to Ti which is further understood by electron density difference. 

From the Mo-d orbital contribution in PDOS, we observe that for Au they exactly overlap each other which implies that MoS$_2$ bandstructure is least affected. This is due to the minimum interaction between graphene-Au interface. However, for Ru and Ti the electronic structure of graphene is highly perturbed, so Mo d-orbital experience a shift w.r.t MoS$_2$-graphene interface. The amount of shift is further examined using the EDD analysis.

\begin{figure*}[!t]
\centering
\includegraphics [scale=0.25]{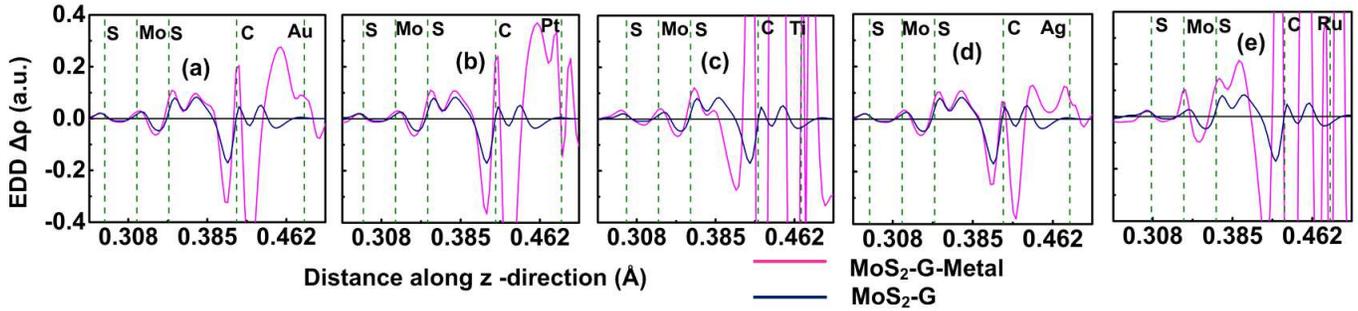}
\caption{ Electron density difference(EDD) for (a) MoS$_2$-G-Au, (b) MoS$_2$-G-Pt, (c) MoS$_2$-G-Ti, (d)MoS$_2$-G-Ag and (e) MoS$_2$-G-Ru, system superimposed with EDD of MoS$_2$-graphene systems. The EDD values on the y-axis are scaled by a factor of 10$^{3}$.}
\label{Figure 6}
\end{figure*}
\subsection{Charge Transfer Investigation}
The shifts observed in CBM/VBM and the amount of perturbations in the orbital contribution is further explored by  evaluating the electron density difference (EDD) averaged along z-direction shown in Figure 5(a)-(e). For MoS$_2$-graphene-metal interface the EDD is calculated as $\Delta\rho=\rho_{MoS_{2}+graphene+metal}-\rho_{MoS_{2}}-\rho_{graphene}-\rho_{metal}$, where $\rho$ is the electron density. The EDD of MoS$_2$-graphene interface is superimposed on the top of MoS$_2$-graphene-metal interface in order develop better understandings of charge transfer occurring at MoS$_2$-graphene-metal interface. While comparing EDD  w.r.t MoS$_2$-graphene interface, the perturbation in Ti and Ru is very high as compared to Au, Ag and Pt at the interface of carbon and interacting sulfur atom of MoS$_2$ since Ti and Ru are chemisorped with graphene.  Both charge accumulation and depletion regions are found at the interface and depending on the metal these vary for the different interfaces. This leads to charge distribution and further dipole formation at the interface which results in band alignment\cite{PhysRevB.64.205310}. We calculate the area under the EDD curve between the nearest sulfur atom and the carbon atoms to analyse the shift in VBM and CBM \cite{my}. Table II shows the value calculated for MoS$_2$-graphene-metal structures for different metals. Positive values imply a higher accumulation region with more chemical interaction at the interface while negative values imply the opposite. Negative values are obtained only for Au and Pt and is higher for Pt. This depicts there is minimal charge transfer from Pt to MoS$_2$-graphene system and surface charge repulsion for both Au and Pt. Hence, we find an increase in n-SBH for Au and Pt systems. On the other hand for Ag, Ti and Ru the area obtained is positive which leads to more accumulation as compared to depletion. The maximum area obtained is for Ru and so we observe a maximum decrease of SBH. \\

\begin{table}
\caption{\label{}Area calculated between the interfacial sulfur atom of MoS$_2$ and carbon atom of graphene for various MoS$_2$-graphene-metal interface.}

\begin{ruledtabular}
\begin{tabular}{l*2{c}}
System & Area Under EDD between C and S atoms \\
MoS$_2$-G & 1.7x10$^{-7}$ \\
MoS$_2$-G-Au & -1.04x10$^{-7}$  \\
MoS$_2$-G-Ag & 3.42x10$^{-7}$  \\
MoS$_2$-G-Ti & 5.62x10$^{-6}$ \\
MoS$_2$-G-Pt & -3x10$^{-7}$ \\
MoS$_2$-G-Ru & 8.1x10$^{-6}$ \\
\end{tabular}
\end{ruledtabular}
\end{table}

\subsection{Work Function calculation and Fermi Level Pinning}
We calculate the work function (WF) of bare and graphene adsorbed metals using the ghost atom technique \cite{WF}, which helps to extend the basis set in the vacuum region. The work function values obtained are 6.05 eV, 5.51 eV, 5.48 eV, 4.82 eV and 4.66 eV for Pt$<$111$>$, Au$<$111$>$, Ru$<$0001$>$, Ag$<$111$>$ and Ti$<$0001$>$ surfaces respectively. They are in near equivalence with the experimental values of those particular surfaces\cite{CRC}. For graphene adsorbed metal the value changes to 4.93 eV, 4.7 eV, 3.47 eV, 4 eV and 4.21 eV for Pt, Au, Ru, Ti and Ag which are consistent with the earlier reports \cite{PhysRevB.79.195425}. The WF of free standing MoS$_2$ and graphene are found to be 5.2 eV and 4.56 eV.\\
 We see a correlation between the change of the SBH of a MoS$_2$-metal system due to graphene insertion and the work function modulation of the metal due to graphene adsorption. The maximum reduction in WF  is seen for Ruthenium and hence we observe large reduction in n-SBH for Ru as well. Figure 6 (a) shows the variation of SBH for MoS$_2$-metal and MoS$_2$-graphene-metal interface w.r.t the  the metal WF. The SBH of MoS$_2$-graphene with a value of 0.65 eV is shown by blue line and acts as the reference to study the alteration of SBH. We see the trend obtained in MoS$_2$-metal is different from the trend obtained MoS$_2$-graphene-metal where the main deviation arises for Ru. Figure 6 (b) features the variation of SBH of MoS$_2$-metal interface and MoS$_2$-graphene-metal interface w.r.t  WF$_{Metal}$-WF$_{MoS2}$ and  WF$_{graphene-metal}$-WF$_{MoS2}$ respectively. The SBH has a linear dependence with a value of 0.61 for MoS$_2$-graphene-metal and 0.31 for MoS$_2$-metal interface. Fitting these characteristics with a linear equation, yields in increase of slope from 0.31 to 0.61 due to graphene insertion. Hence it could be inferred that graphene insertion helps to de-pin the Fermi-level partially in a MoS$_2$-metal interface.  \\

In the above discussion we explain the underlying mechanism of SBH change of the MoS$_2$-graphene interface, when a metal slab is placed beneath the graphene. However it is difficult to conceive similar explanation for the change of SBH with respect to the MoS$_2$-metal interface. This is because the contact nature of the MoS$_2$-graphene-metal system is dictated by the graphene-metal interaction which is very different from the nature of interaction observed in MoS$_2$-metal interface.
 
\begin{figure}[!t]
\centering
\includegraphics [scale=0.38]{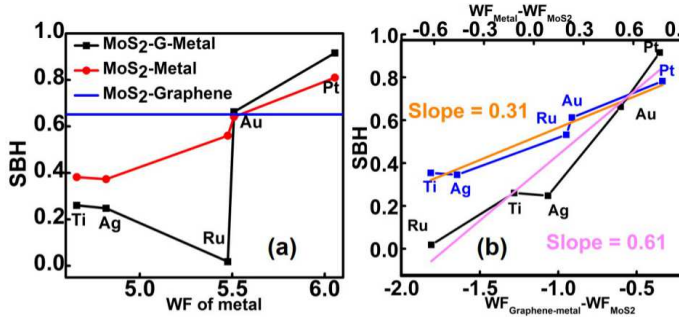}
\caption{(a) Plot highlighting comparison of SBH of MoS$_2$-metal and MoS$_2$-graphene-metal interface w.r.t the work function of various metals. The SBH for MoS$_2$ obtained with MoS$_2$-graphene system is shown by blue line depicting the increase and decrease of SBH of metals in MoS$_2$-graphene-metal systems. (b) Dependence of MoS$_2$-metal SBH (blue line) versus  WF$_{Metal}$-WF$_{MoS2}$ and  MoS$_2$-graphene-metal SBH (black line) versus  WF$_{graphene-metal}$-WF$_{MoS2}$ for all the metals. The orange and pink lines shows the linear fitting of the two curves. }
\label{Figure 8}
\end{figure}

\section{\label{sec:level1}Conclusion}
 In conclusion, we have conducted rigorous DFT calculation to investigate the charge transfer from metal to MoS$_2$  in a graphene inserted MoS$_2$-metal contact involving five different metals (Ti, Ag, Ru, Au and Pt). Graphene acts a perfect buffer separating MoS$_2$ from metal and thus retains the band gap nature with minimal interface states. Different metals showed varying behavior and inserting graphene in a metal MoS$_2$ contact does not assure a SBH reduction. An increase in SBH is observed for Au and Pt while a decrease for Ag, Ti and an Ohmic nature is found for Ru. A large fluctuation in the band alignments is due to the interface charge transfer which further leads to the dipole formation. The variation observed in SBH is highly dependent on the nature of the graphene-metal interface. These findings can lead to further design of high performance transistors using heterostructures as contacts.

\begin{acknowledgements}

The work was supported by the Science Engineering and Research Board, Department of Science and Technology, Government of India, under Grant SR/S3/EECE/0151/2012.

\end{acknowledgements}

\bibliography{Manuscript}

\end{document}